\documentstyle[preprint,aps,prb]{revtex}

\newcommand{\BI}{\mathrm{Bi_2Sr_2CaCu_2O_{8+\delta}}}
\begin{document}
\draft

\title{Superconducting Plasma Excitation at Microwave Frequencies in Parallel Magnetic Fields in $\mathrm{\mathbf{Bi_2Sr_2CaCu_2O_{8+\delta}}}$}

\author{K. Kadowaki, T. Wada and I. Kakeya}
\address{Institute of Materials Science, The University of Tsukuba, Tsukuba, Ibaraki 305-8573, Japan, and \\ CREST, Japan Science and Technology Corporation (JST), Japan}

\date{\today}
\maketitle
\begin{abstract}

Josephson plasma resonance has been studied in a wide microwave frequency range between 10 and 52 GHz in a magnetic field parallel to the $ab$-plane in under-doped $\BI$.  Above about 30 GHz two resonance modes were observed: one (LT mode) appears at low temperatures and another (HT mode) at higher temperatures, leaving a temperature gap between two regions.  These two resonance modes exhibit a sharp contrast each other both on temperture and magnetic field dependences and show distinct characters different entirely from the $c$-axis Josephson plasma resonance.  From temperature and field scan experiments at various frequencies it is suggested that the LT mode can be attributed to the coupled Josephson plasma mode with Josephson vortices, while the HT mode is a new plasma mode associated possibly with the periodic array of Josephson vortices.
\end{abstract}


\narrowtext

\section{introduction}
It has been well established that a sharp microwave absorption can be observed in high temperature superconductors with a strong superconducting anisotropy such as $\BI$\cite{Tsui1}.  The phenomenon itself manifests the existence of plasma oscillation of the weak superconducting current through a stuck of seriesly connected intrinsic Josephson junctions associated with the layer structure of the material.  The Josephson plasma frequecny can well be described as\cite{Fertig}
\begin{eqnarray}
\omega_p(0,T)=\frac{c}{\sqrt{\epsilon}\lambda_c(T)}=\frac{c}{\sqrt{\epsilon}\gamma\lambda_{ab}(T)},
\end{eqnarray}
where $\epsilon$ is the dielectric constant at high frequencies of the mateiral, $\gamma$ is the anisotropy parameter and $\lambda_{ab,c}$(=$(c\phi_0/8\pi^2\{j_{ab},j_c\} s)^{1/2}$=$(hc^2/16\pi^2e\{j_{ab},j_c\} s)^{1/2}$) denotes the penetration depths for the $ab$-plane and for the $c$-axis.  $j_{ab}$ and $j_c$(=$j_J$) mean the critical current for the $ab$-plane and the Josephson critical current for the $c$-axis, respectively, $s$ the interlayer distance between two adjacent superconducting layers and $\phi_0$=$hc/2e$ the flux quantum.  This is essentially identical with the case of single junction worked out by Anderson\cite{Anderson} and Josephson\cite{Josephson}, and later Lawrence and Doniach\cite{Lawrence}.

In magnetic fields above $H_{c1}$ the vortices penetrate through the superconducting material and induce the relative shift of the gauge invariant phase $\varphi$ of the superconducting order parameter between two adjacent layers.  The phase shift is apparently caused by the Josephson current flow which depends on how vortices are arranged between two adjacent layers.  When magnetic field is applied perpendicularly to the layers ($c$-axis) the plasma frequency $\omega_p(H,T)$ can be expressed as\cite{Bulaevskii1,Koshelev}
\begin{eqnarray}
\omega_p^2(H,T)=\omega_p^2(0,T)\langle\cos\varphi_{n,n+1}\rangle(H,T),
\end{eqnarray}
where 
\begin{eqnarray}
\varphi_{n,n+1}=\varphi_{n+1}-\varphi_n-\frac{2\pi}{\phi_0}\int_n^{n+1}A_zdz
\end{eqnarray}
means the gauge invariant phase difference between $n$-th and $n+1$-th layers, $A_z$ the vector potential and $\langle\cdot\cdot\cdot\rangle$ stands for the thermal and positional average.  Recently, there have been a great deal of work on the Josephson plasma resonanace\cite{Bulaevskii2,Bulaevskii3,Ohashi1,Ohashi2,Koyama,Uchida}, especially on the vortex state of $\BI$\cite{Matsuda1,Kadowaki1,Kadowaki2,Kadowaki3,Kakeya} and all experimental results have been explained reasonably well along this line.

In our previous publications we have reported preliminary results of the angular dependence of Josephson plasma respnance\cite{Kadowaki1,Kadowaki2,Matsuda2}: a sharp dip structure of the resonance field was observed, when the magnetic field orientation comes within $\pm$1.0 $^\circ$ from the $ab$-plane.  This implies that the anisotropic Ginzburg-Landau(GL) scaling law is violated in this narrow angular region.  Tsui \textit{et. al.}\cite{Tsui2} reported a similar sharp dip behavior within $\pm5 ^{\circ}$, however, the angular dependence seems to follow a different function, resulting in the contradictory results.  When magnetic field is applied exactly parallel to the layers (within accuracy of $\pm$0.2 $^{\circ}$ from the $ab$-plane), no resonance has been reported previously\cite{Matsuda2}.  Although these experimental results have been argued by comparing currently available theories\cite{Koshelev,Bulaevskii4,Bulaevskii5,Bulaevskii6}, agreement is far from satisfactory.

We have made continuous effort to obtain full description of the Josephson plasma resonance in a field near the $ab$-plane and found entirely new results, which have never been explored before.  We discovered two resonance modes appearing in two different temperature regemes: the LT mode appearing at low temperatures and the HT mode at high temperatues.  These two modes are separated by the temperature gap, which depends strongly on the microwave frequency.  We believe that the LT mode is the Jpsephson plasma resonance coupled with the Josephson vortex kattice system, while the HT mode resembles the plasma mode modified by the periodic array of Josephson vortices.

Here we will give the comprehensive experimental results with emphasis on new discoveries so far we have obtained on the issue of the Josephson plasma in magnetic fields parallel to the $ab$-plane.

\section{experiments}
Josephson plasma resonance experiments have been performed at various fixed frequencies between 10 and 52 GHz either by sweeping temperature in a fixed magnetic field or by sweeping magnetic field at a fixed temperature.  The signal generator HP83650B (Hewlett Packard Co.), the klystrons (Okaya Denki Co. Ltd.) and the gunn oscillators (Quinstar Co.) have been used depending on the frequencies.  The cavity resonator coupled with the magic T bridge balance circuit was installed in a variable temperature insert and was rotated by a fine goniometer system in a 7 T split magnet.  The precise field direction parallel to the $ab$-plane was determined from the symmetry of the Josephson plasma resonance position with respect to the angle near the $ab$-plane.  The error of the angular setting can be as small as 0.01 $^{\circ}$.  Temperature was controled better than 0.01 K during the magnetic field scan measurement.

The $\BI$ sample used in the present study was grown by the improved traveling solvent floating zone (TSFZ) method.  After annealing in a reduced atmosphere an under-doped $\BI$ was obtained with $T_c$=70.2 K and the transition width of $\Delta T_c\lesssim$1.5 K.  The sample was cut into a rectangle with the dimension of 0.7 mm $\times$ 0.8 mm $\times$ 20 $\mu$m and was set at a position of a maximum microwave electric field in the cavity resonator of the $\mathrm{TE_{102}}$ mode.

\section{results and discussion}
The typical Josephson plasma resonance curves in $\BI$ in parallel magnetic fields is demonstrated in Fig. 1 as an example.  Here, the direct record of the microwave absorption at 25.5 GHz and 34.30 GHz measured by the field sweep mode at various temperatures are presented.  First of all, there are obviously two Josephson plasma modes(LT and HT modes) in two temperature regions separated by a temperature gap which strongly depends on the microwave frequency.  The gap can only be observable between 20 and 32 GHz, which will be shown later more clearly in Fig. 4.  Since a larger gap in temperature is found at higher frequencies and the LT mode shifts quickly to lower temperatures as the frequency is icreased, the LT mode cannot be observed above about 35 GHz in this under-doped sample. As will clearly be seen later in Fig. 4, both LT and HT resonances merge together below 10 GHz.

Secondly, the HT mode at a fixed temperature has two resonances as a function of magnetic fields.  This feature has never been observed in the previous studies both in parallel to the c-axis and in tilted fields.  Furthermore, the higher field resonance of the HT mode once tends to decrease but above about 0.8 kOe it turns to shift to a higher temperature side with increasing field.  It does not shift back to zero field, exhibiting a strong contrast with the Josephson plasma resonance for the case in fields parallel to the c-axis.  It is surprizing that the higher field resonance of the HT mode is stable even in temperatures where the zero field plasma resonance has already disappeared.  This fact strongly suggests that the higher field resonance of the HT mode has higher plasma energy than $\omega_{p}(0,T)$ in zero field.  Acording to Bulaevskii \textit{et. al.}\cite{Bulaevskii4,Bulaevskii5,Bulaevskii6} on the basis of single junction\cite{Lebwohl,Fetter}, it is only possible for the plasma oscillation mode coupled with the Josephson vortex system to possess higher energies in a magnetic field than $\omega_p(0)$ in zero filed.  Fom this reason, we attribute the HT mode to the new Josephson plasma reosnance mode associated with the periodic array of the Josephson vortices.  This mode perhaps corresponds to the plasma ``\textit{oscillations}'' in the case of a single Josephson junction\cite{Lebwohl,Fetter}.

On the contrary to this HT mode, the LT mode shows a considerable hysteresis effect of the resonance position, the intensities as well as the resonance width on magnetic field sweeping conditions as seen in Fig. 1.  This hysteresis effect becomes large especially below 25 K, suggesting existence of the different pinning state at low temperatures.  This implies that the resonance occurs in non-equilibrium states of Josephson vortices during sweeping magnetic field.

In order to avoid the hysterisis behavior ovserved in the field sweep measurements the tempreature sweep mode is operated at a fixed magnetic field, whcih is applied above $T_c$.  Figure 2 shows an example of the case for 25.5 and 30.1 GHz.  The HT mode shifts from zero field resonance towards low temperatures in a finite field below about 0.8 kOe, but above this field it is surprising that it turns over towards higher temperatures.  Although the intensity of the resonance after the turn-over becomes fainter and fainter, it is clearly seen that the resonance is sharp and it goes beyond the temperature where the zero field plasma resonance was observed.  There is no resonance observed in-between the HT and LT resonances in this temperature sweep mode.  As seen in Fig. 2, the LT mode is broad in temperature and shifts to lower temperatures in higher magnetic fields.

These overall behavior at 30.1 GHz is summerized in Fig. 3, where the results obtained from both field and temperature sweep measurments are plotted together.  In the temperature gap region between LT and HT modes a faint absorption observed only in the field sweep measurements were also included.  In the previous study\cite{Matsuda2} sudden disappearance as well as entire absence of the plasma resonance were reported.  However, it is apparent from the present results that the reason of disappearance of the resonance is due to the temperature gap existing in-between LT and HT modes, which expands quickly as a function of frequency as disclosed in Fig. 3.  Therefore, at certain frequencies and temperature regions the resonance cannot be observed.  We further comment on the resonance below about 25 K. It was not possible to find any meaningful absorption in the field sweep mode measurement because of broadening of the absorption as well as quick shift of the resonance field to zero.  Therefore, we could not confirm the results by Tsui \textit{et. al.}\cite{Tsui2}

In Fig. 4, the resonance field-temperature diagram similar to Fig. 3 is presented at seven different frequencies 9.80, 18.8, 25.5, 30.1, 34.5, 44.2 and 52.3 GHz.  As seen in Fig. 4 the gap between LT and HT modes rapidly grows at higher frequencies.  The LT resonance is no longer seen above about 30 GHz due to this fact.  On the contrary to this, the HT mode tends to converge at about 3 kOe after the turn-over and disappears in higher fields.  Although the Josephson plasma resonance at zero field nicely obeys the temperature dependence as described previously\cite{Kadowaki4}, the fact shown above indicates that the higher field resonance of the HT resonance should possess higher plasma frequency than the zero field plasma frequency.

The frequency-field diagram is summerized in Fig. 5.  It is clearly seen the existence of two resonance modes: the LT mode frequency is suppressed by magnetic field, whereas the HT mode turns over at about 0.8 kOe then it goes up to higher frequencies.  We note that in the low field region below 0.5 kOe the LT mode tends to shift to zero frequencies as indicated by the extrapolated lines in Fig. 5, indicating a gapless feature of the LT mode.  Furthermore, the fact that this LT mode is only observable in fields suggests that the LT mode is related to the existence of the Josephson vortices in the system.  At higher fields the LT mode monotonically decreases with increasing magnetic field.  On the other hand, the field dependence of the HT mode cannot be understood by the presently available models for the Josephson plasma resonance in which a reduction of the Josephson current in magnetic fields simply is considered in reducing the resonance frequencies.

In order to better understand the unusual behabior of HT mode described above in Fig. 5, we first note that the Josephson vortex state in a parallel field is not yet well established.  In general, it has been believed that several phases with complicated vortex structures may exit.  A successive first order phase transitions were predicted by Bulaevskii\cite{Bulaevskii7}, whereas occurrence of more complicated phases such as a floating and a waving solids\cite{Ikeda}, and smectic and supersolid\cite{Balents} are also theoretically proposed.

Recently, we have studied the transport properties of the vortex state in a parallel magnetic field in detail and found unexpectedly rich phenomena\cite{Mirkovic}.  One of them is a continuous phase transition at $H_{SM}$ from Josephson vortex liquid state to Josephson vortex solid state in optimally doped $\BI$.  Non-linear behavior of the resistivity is also significant and characteristic in this region.  Since the doping level of the sample used here is under-doped and different from the one, it seems that the resonance field $H_{res}$ in all temperatures are much smaller than $H_0$, where all layeres are filled by the Josephson vortices.  This means that the Josephson plasma resonance observed here are perhaps all in the Josephson votex solid phase ($H_{res}<H_0<H_{SM}$), which is not fully explored yet.

Considering such a complicated phases in parallel magnetic fields it is furthermore expected that the plasma dispersion relation can be modified strongly from the one for the case of fields parallel to the $c$-axis by the periodic arrangement of Josephson vortices in-between superconducting layers in such a way that due to the periodic array of Josephson vortices it can be reduced at the zone boundary, resulting in the multi-band scheme similar to the electron bands in solids.  In our experiemental conditions the plasma excited by microwave frequencies must be $k\sim$ 0.  Nevertheless, it may be possible to excite plasma with finite $\pm \mbox{\boldmath{$k$}}$, leaving $\mp \mbox{\boldmath{$k$}}$ behind with an aid of a reciprocal vector $\mbox{\boldmath{$G$}}/2$=$\mbox{\boldmath{$k$}}$ of the Josephson lattice.  If this mechanism works efficiently, the higher field resonance may correspond to the higher energy modes associated with the occurence of multiple plasma bands due to the periodic array of the Josephson vorices.

\section{conclusion}
We have studied Josephson plasma resonance in under-doped single crystalline $\BI$ in parallel magnetic field configuration in detail.  We discovered for the first time two well-separated resonance branches above 20 GHz in two different temperture regions.  Both resonances behave the peculiar temperature, magnetic field and frequency dependeces unlike the resonance for the $c$-axis.  In our experimental condition that the microwave $\mbox{\boldmath{$E$}}$-vector is perpendicular to the $ab$-plane, which selectively excite the longitudinal plasma mode\cite{Kakeya}, we point out that there must be a mechanism which converts the longitudinal plasma to the transverse plasma modes.  It is proposed that the momentum transfer may be possible if the Josephson vortex lattice vibrational (translational) motion are involved in the fundamental processes.  We infer that the HT mode may be the optical-like plasma mode modified by the motion of the periodic arrangement of Josephson vortices, whereas the LT mode may be the acoustic-like mode coupled with the Josephson vortex lattice vibration modes.  Further studies are reqired in future in order to confirm the above mentioned conjectures.

\begin{figure}[btp]
\begin{center}
\leavevmode
\caption{Josephson plasma resonance absorption curves are shown in a magnetic field parallel to the $ab$-plane using the field sweep mode at 25.5 GHz and 30.1 GHz at fixed various temperatures.  The sharp spike-like absorption at 9.093 kOe is due to the electron spin resonance of DPPH($\alpha$, $\alpha$'-diphenyl-$\beta$-picryl hydrazyl) as a field marker.  The magnetic field was swept at a constant speed of about 5 Oe/sec.}
\label{figure1}
\end{center}
\end{figure}

\begin{figure}[btp]
\begin{center}
\leavevmode
\caption{Josephson plasma resonance absorption curves in a magnetic field parallel to the $ab$-plance using the temperature sweep mode at 25.5 GHz and 30.1 GHz at various fixed fields.  The temperature is swept at a constant speed of 1.0 K/min.  No hysteresis was observed in this mode.}
\label{figure2}
\end{center}
\end{figure}

\begin{figure}[btp]
\begin{center}
\leavevmode
\caption{ The Josephson plasma resonance field plotted as a function of temperature obtained by both temperature and field sweep modes at 30.1 GHz.  In the temperature gap region the weak resonance can only be observable in magnetic field sweep mode.}
\label{figure3}
\end{center}
\end{figure}

\begin{figure}[btp]
\begin{center}
\leavevmode
\caption{The Josephson plasma resonance field as a function of temperature at various frequencies of 9.80, 18.8, 25.5, 30.1, 34.5, 44.2 and 52.3 GHz.}
\label{figure4}
\end{center}
\end{figure}

\begin{figure}[btp]
\begin{center}
\leavevmode
\caption{The Josephson plasma resonance frequency as a function of the resoannce field.  $\omega_p(0)\simeq60.0$ GHz is the plasma gap frequency in zero field estimated form the temperature dependnece of zero field resonance\cite{Kadowaki4}.  The filled symbols correspond to the HT mode, whereas the open symbols correspond to the LT mode as indicated in the figure.}
\label{figure5}
\end{center}
\end{figure}

\begin{references}
\bibitem{Tsui1} Ophelia K. C. Tsui, N. P. Ong, Y. Matsuda, Y. F. Yan and J. B. Peterson, Phys. Rev. Lett. \textbf{73} (1994) 724.
\bibitem{Fertig} H. A. Fertig and S. Das Sarma, Phys. Rev. Lett. \textbf{65} (1990) 1482, H. A. Fertig and S. Das Sarma, Phys. Rev. \textbf{B44} (1990) 4480.
\bibitem{Anderson} P. W. Anderson, \textit{``Lectures on the Many-Body Problem''} Vol. 2, edited by E. R. Caianiello, 1994, Academic Press, p.112-135.
\bibitem{Josephson} B. D. Josephson, Adv. Phys. \textbf{14} (1965) 419.
\bibitem{Lawrence} W. E. Lawrence and S. Doniach, ``\textit{Proc. 12th Int. Conf. Low Temp. Phys.''}, edited by E. Kanda, 1970, Tokyo Shokabo, p361.
\bibitem{Bulaevskii1} L. N. Bulaevskii, M. P. Maley and M. Tachiki, Phys. Rev. Lett. \textbf{74} (1995) 801.
\bibitem{Koshelev} A. E. Koshelev, Phys. Rev. Lett. \textbf{77} (1996) 3901.
\bibitem{Bulaevskii2} L. N. Bulaevskii, V. L. Pokrovsky and M. P. Maley, Phys. Rev. Lett. \textbf{76} (1996) 1719.
\bibitem{Bulaevskii3} L. N. Bulaevskii, D. Dom\'inguez, M. P. Maley, A. R. Bishop, Ophelia K. C. Tsui and N. P. Ong, Phys. Rev. \textbf{B54} (1996) 7521.
\bibitem{Ohashi1} Y. Ohashi and S. Takada, J. Phys. Soc. Jpn. \textbf{66} (1997) 2437.
\bibitem{Ohashi2} Y. Ohashi and S. Takada, J. Phys. Soc. Jpn. \textbf{67} (1998) 551.
\bibitem{Koyama} T. Koyama and M. Tachiki, Phys. Rev. \textbf{B54} (1996) 16183
\bibitem{Uchida} S. Uchida and K. Tamasaku, Physica \textbf{C293} (1997) 1, (see also ``\textit{Proc. 1st RIEC Int. Symp. Intrinsic Josephson Effects and THz Plasma Oscillations in High $T_c$ Superconductors}'', edited by M. Tachiki and T. Yamashita, Sendai, Miyagi, Japan, Feburary 23-25, 1997.)
\bibitem{Matsuda1} Y. Matsuda, M. B. Gaifullin, K. Kumagai, K. Kadowaki and T. Mochiku, Phys. Rev. Lett. \textbf{75} (1995) 4512, Y. Matsuda, M. B. Gaifullin, K. Kumagai, M. Kosugi and K. Hirata, Phys. Rev. Lett. \textbf{78} (1997) 1972.
\bibitem{Kadowaki1} K. Kadowaki, I. Kakeya, K. Kindo, S. Takahashi, T. Koyama and M. Tachiki, Physica \textbf{C293} (1997) 130.
\bibitem{Kadowaki2} K. Kadowaki, I. Kakeya and T. Mochiku, Physica \textbf{B239} (1997) 123.
\bibitem{Kadowaki3} K. Kadowaki, I. Kakeya, M. B. Gaifullin, T. Mochiku, S. Takahashi, T. Koyama and M. Tachiki, Phys. Rev. \textbf{B56} (1997) 5617.
\bibitem{Kakeya} I. Kakeya, K. Kindo, K. Kadowaki, S. Takahashi and T. Mochiku, Phys. Rev. \textbf{B57} (1998) 3108.
\bibitem{Matsuda2} Y. Matsuda, M. B. Gaifullin, K. Kumagai, K. Kadowaki, T. Mochiku and K. Hirata, Phys. Rev. \textbf{B55} (1997) R8685.
\bibitem{Tsui2} Ophelia K. C. Tsui, N. P. Ong and J. B. Peterson, Phys. Rev. Lett. \textbf{76} (1996) 819.
\bibitem{Bulaevskii4} L. N. Bulaevskii, D. D. Dominguez, M. P. Maley, A. R. Bishop and B. I. Ivlev, Phys. Rev. \textbf{B53} (1996) 14601.
\bibitem{Bulaevskii5} L. N. Bulaevskii, M. P. Maley, H. Safar and D. Dominguez, Phys. Rev. \textbf{B53} (1996) 6634.
\bibitem{Bulaevskii6} L. N. Buklaevskii, D. Dominguez, M. P. Maley and A.R. Bishop, Phys. Rev. \textbf{B55} (1997) 8482.
\bibitem{Lebwohl} P. Lebwohl and M. J. Stephen, Phys. Rev. \textbf{163} (1967) 376.
\bibitem{Fetter} A. L. Fetter and M. J. Stephenm Phys. Rev. \textbf{168} (1968) 475.
\bibitem{Kadowaki4} K. Kadowaki, I. Kakeya, T. Wakabayashi, R. Nakamura and S. Takahashi, J. Mod. Phys. \textbf{B14} (2000) 547.
\bibitem{Bulaevskii7} L. N. Bulaevskii and J. R. Clem, Phys. Rev. \textbf{B44} (1991) 10234.
\bibitem{Ikeda} R. Ikeda and K. Isotani, J. Phys. Soc. Jpn. \textbf{68} (1999) 599.
\bibitem{Balents} L. Balents and D. R. Nelson, Phys. Rev. \textbf{B52} (1995) 12951.
\bibitem{Mirkovic} J. Mirkovi\'c, S. Savel'ev, E. Sugahara and K. Kadowaki, to be published in Phys. Rev. Lett.
\end{references}
\end{document}